\documentclass[aps,showpacs,10pt,pra,amsmath,amssymb,superscriptaddress,twocolumn]{revtex4-1}

\usepackage{graphicx}
\usepackage{amsmath}
\usepackage{hyperref}
\usepackage{dcolumn}
\usepackage{soul}
\usepackage{color}
\usepackage{bm}
\usepackage{braket}

\begin{document}

\title{Coherent population oscillations and an effective spin-exchange interaction
in a $\mathcal{PT}$ symmetric polariton mixture}

\author{P.A. Kalozoumis}
\affiliation{ Materials Science Department, School of Natural Sciences, University of Patras, GR-26504 Patras, Greece}
\affiliation{Hellenic American University, 436 Amherst st, Nashua, NH 0306 USA}
\affiliation{Institute of Electronic Structure and Laser, FORTH, GR-71110 Heraklion, Crete, Greece}

\author{G. M. Nikolopoulos}
\affiliation{Institute of Electronic Structure and Laser, FORTH, GR-71110 Heraklion, Crete, Greece}

\author{D. Petrosyan}
\affiliation{Institute of Electronic Structure and Laser, FORTH, GR-71110 Heraklion, Crete, Greece}
\affiliation{A. Alikhanyan National Laboratory, 0036 Yerevan, Armenia}


\begin{abstract}
We study a two-species mixture of exciton-polaritons with 
self- and cross-interaction nonlinearities in a double well structure, 
in the presence of relaxation and continuous pumping. 
We identify the conditions that render the system 
parity-time ($\mathcal{PT}$) symmetric, and investigate its dynamic and 
static properties. We show that the system can exhibit long-term coherent 
oscillations of populations of the two polaritonic components between the 
two potential wells, and can simulate the dynamics of a pair of spin-1/2 
particles (qubits) in the presence of exchange interaction.
\end{abstract}

\maketitle

\section{Introduction}
\label{sec:Int}

Exciton-polaritons are elementary excitations of semiconductor 
microcavities and constitute hybrid quasi-particles of strongly 
coupled light (cavity photons) and matter (quantum well excitons) excitations 
\cite{DengRevModPhys2010,CarusottoRevModPhys2010}, retaining the properties 
of both constituents. The excitons instill effective interactions inducing 
polariton nonlinearities, whereas the small effective mass of photons
enables Bose-Einstein condensation (BEC) of polaritons 
at high temperatures~\cite{YamaguchiPRL2013,SchneiderNature2013}.
First experimental observations of exciton-polariton BECs have been 
reported more than a decade ago 
\cite{KasprzakNature2006,BaliliScience2007}. 
Since then, the remarkable properties of the exciton-polariton systems,
combined with their high condensation temperatures \cite{LerarioNatPhys2017},
have motivated much of the research activity in this field. 
BECs of exciton polaritons have been studied in various geometries, 
e.g., parabolic traps \cite{WoutersPRB2008,TralleroPRB2014}, 
double-well potentials \cite{WoutersPRL2007,LagoudakisPRL2010,RahmaniSciRep2016,AbbarchiNatPhys2013},
triple wells \cite{ZhangOQE2017}, and 1D and 2D polariton lattices
\cite{AmoPhysique2016,KlembtAPL2017,WhittakerPRL2018,OhadiPRB2018,PanPRB2019}. 
In contrast to relatively stable atomic BECs, the polariton BECs 
are open quantum systems; they are inherently strongly dissipative 
and require continuous pumping, e.g., by external laser fields 
\cite{BallariniNatCom2013}.

The dynamics of dissipative systems can be rendered pseudo-Hermitian 
in the parity-time ($\mathcal{PT}$) symmetric setup 
\cite{Bender1999,Bender2007,ElGanainy2018}, where the gain and loss 
are exactly balanced in a complex potential with the reflection symmetric 
real part (energy) and reflection antisymmetric imaginary part (gain/loss). 
Systems with $\mathcal{PT}$-symmetry have recently attracted much interest
in various branches of physics, 
extending from quantum mechanics \cite{Bender2002} 
and field theory \cite{Bender2004} to optics \cite{Ganainy2007, Musslimani2008a,Peng2014} 
and acoustics \cite{Zhu2014,Fleury2015}. 
Since the experimental realization of the spontaneous $\mathcal{PT}$ symmetry 
breaking \cite{Guo2009,Ruter2010}, there has been an enormous progress 
in the field and a multitude of interesting phenomena have been observed, 
such as, e.g., power oscillations \cite{Ruter2010}, 
double refraction \cite{Makris2010}, and non-reciprocal diffraction \cite{Makris2008}. 
The presence of nonlinearity renders $\mathcal{PT}$ symmetric 
systems even more remarkable, permitting, for example,
unidirectional \cite{Ramezani2010} and asymmetric \cite{Ambroise2012,Yang2015} 
wave propagation in discrete and continuous structures 
\cite{Kottos2010,Li2011,Li2013,Zezyulin2012}. 

Despite the plethora of different settings, the study of $\mathcal{PT}$ 
symmetry in polariton structures is still in its infancy and most works 
address this issue in the single species context
\cite{LienPRB2015,GaoNature2015}. 
A natural next step is to consider polariton mixtures
\cite{SatijaPRA2009,ChestnovSciRep2016,Ohadi2017} in a $\mathcal{PT}$ symmetric
setting where the losses of the polaritons -- due to the semiconductor 
exciton recombination and photon escape for the microcavity -- are 
compensated by gain from the external laser pumping of the exciton population. 
Here we study a two-species polariton mixture with self- and cross-interaction 
nonlinearities in a double well structure. We identify the necessary condition 
imposed on the pumping rate of the reservoirs and their losses to render
the polariton mixture $\mathcal{PT}$ symmetric. We construct the corresponding 
model, analyze its static and dynamic properties, and show that nearly
perfect Rabi-like oscillations of the two polariton components between the
two wells can be observed in this system in the presence of moderate self- 
and cross-interactions. This system can then formally be mapped onto 
a system of two qubits, or spin-$1/2$ particles, coupled via
exchange ($XY$) interaction. We calculate the fidelity of an effective
 \textsc{swap} gate and quantum state transfer that the system can simulate,
despite being an essentially classical system of coupled BECs. 
Our results can thus have interesting implications for 
simulations of spin models with polariton lattices.

\section{The exciton-polariton system}
\label{II}

\begin{figure}[t]
\begin{center}
\includegraphics[width=\columnwidth]{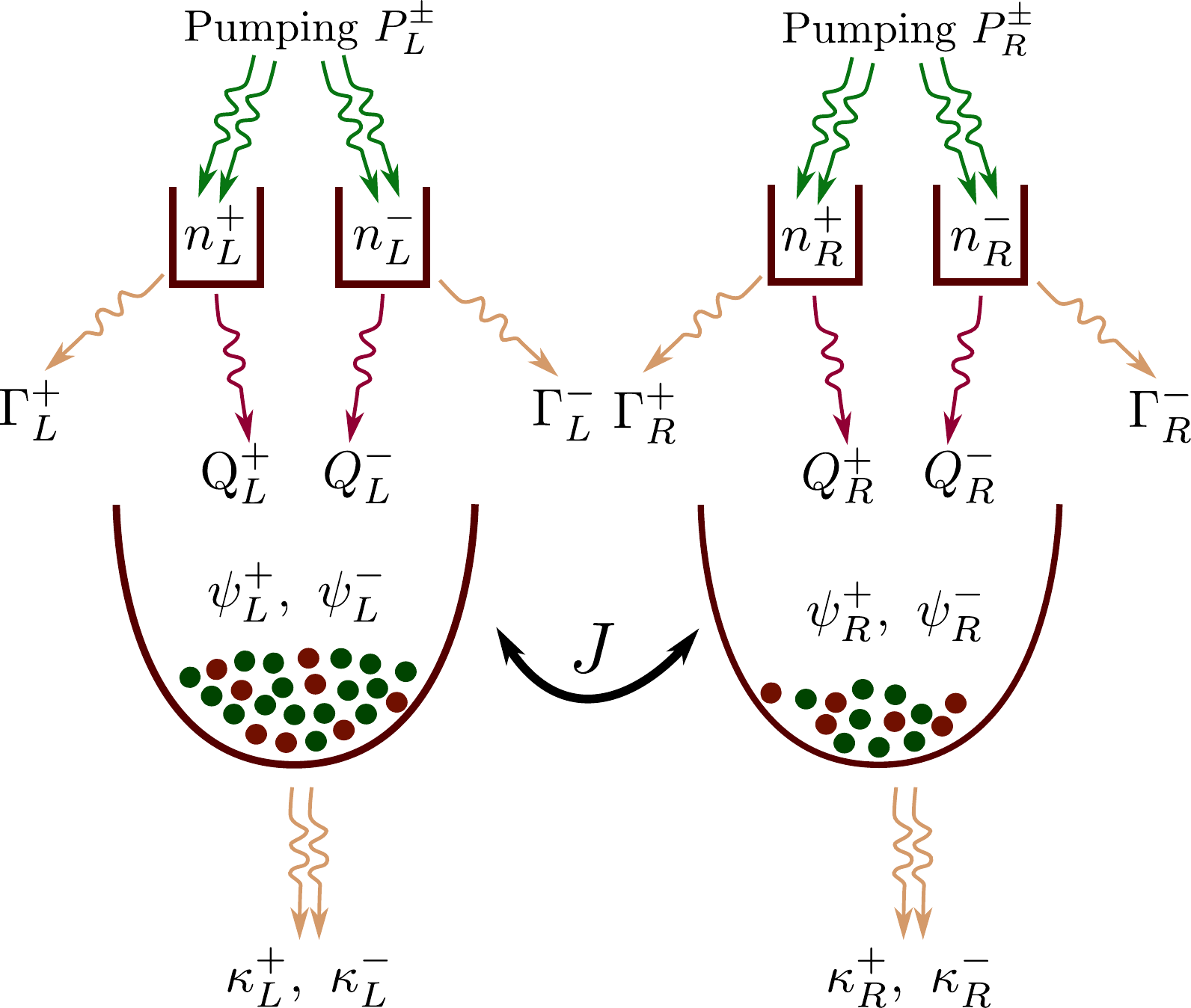}
\end{center}
\caption{Schematic representation of a coupled double well potential 
with a binary mixture of exciton-polaritons $\psi_{L,R}^{\pm}$. 
Four exciton reservoirs $n_{L,R}^{\pm}$ replenish each polariton 
species $(+)$, $(-)$ in each well ($L$), ($R$), respectively. 
By balancing the polariton loss $\kappa_{L,R}^{\pm}$ and gain 
$Q_{L,R}^{\pm}$ via appropriate reservoir pumping $P_{L,R}^{\pm}$, 
we can render the system $\mathcal{PT}$ symmetric.}
\label{fig:fig1}
\end{figure}

We consider a mixture of two species of semiconductor cavity  
exciton polaritons in a double-well potential, 
as shown schematically in Fig.~\ref{fig:fig1}. 
Each of the species $(+)$ or $(-)$ represents a polariton composed 
of a circularly right- or left-polarized cavity photon coupled 
to a semiconductor exciton transition ($\sigma_+$ or $\sigma_-$) with 
the corresponding change of the magnetic quantum number $\Delta m = +1$ or $-1$. 
Starting with the mean-field Gross-Pitaevskii equation for 
the exciton-polariton BEC in a tight-binding double-well potential, 
we follow the standard procedure \cite{RaghavanPRA1999} to derive 
the discrete nonlinear Schr\"odinger equation for the four-component 
system represented by the wavefunctions $\psi_{L}^{\pm}$ and $\psi_{R}^{\pm}$
of the $(\pm)$ species in the left ($L$) and right ($R$) wells,
respectively,
\begin{subequations} 
\label{eqs:EPs}
\begin{eqnarray}
i \hbar \dot{\psi}_{L}^{\pm} &=& 
\varepsilon_{L}^{\pm} \psi_{L}^{\pm} + g_s|\psi_{L}^{\pm}|^{2}\psi_{L}^{\pm} 
+ g_c|\psi_{L}^{\mp}|^{2}\psi_{L}^{\pm} - J \psi_{R}^{\pm} 
\nonumber \\ & & 
 + i \frac{\hbar}{2} [Q_{L}^{\pm}(n_{L}^{\pm}) - \kappa_{L}^{\pm} ] \psi_{L}^{\pm} , \\
i \hbar \dot{\psi}_{R}^{\pm} &=& 
\varepsilon_{R}^{\pm} \psi_{R}^{\pm} + g_s|\psi_{R}^{\pm}|^{2}\psi_{R}^{\pm} 
+ g_c |\psi_{R}^{\mp}|^{2}\psi_{R}^{\pm} - J \psi_{L}^{\pm} 
\nonumber \\ & & 
 + i\frac{\hbar}{2} [Q_{R}^{\pm}(n_{R}^{\pm}) - \kappa_{R}^{\pm} ] \psi_{R}^{\pm} .
\end{eqnarray}
\end{subequations}
Here $\varepsilon_{L,R}^{\pm}$ are the single-particle energies in each well, 
$g_{s}$ and $g_{c}$ are the self- and cross-interaction strengths of the
polaritons, and $J$ is the Josephson tunnel-coupling between the two wells, 
assumed the same for both species. The complex wavefunctions can be written 
as $\psi_{L,R}^{\pm} = \sqrt{N_{L,R}^{\pm}} \, e^{i \phi_{L,R}^{\pm}}$, where
$N_{L,R}^{\pm} \equiv |\psi_{L,R}^{\pm}|^2$ are the polariton populations
and $\phi_{L,R}^{\pm}$ are the phases. 
In contrast to an atomic BEC \cite{RaghavanPRA1999}, the polariton BEC 
is a non-conservative system, with the decay $\kappa^{\pm}_{R,L}$ 
due the exciton recombination losses and cavity photon losses.
The polariton population is continuously replenished via coupling to 
exciton reservoirs with rates $Q_{R}^{\pm}(n_{R}^{\pm})$~\cite{WoutersPRL2007}.  
In turn, the exciton populations in each reservoir $n_{R,L}^{\pm}$ 
obey the equations
\begin{subequations} 
\label{eqs:reservoirs}
\begin{eqnarray}
\dot{n}_{L}^{\pm} &=& P_{L}^{\pm} - \Gamma_{L}^{\pm} n_{L}^{\pm} - 
Q_{L}^{\pm}(n_{L}^{\pm}) N_{L}^{\pm} , \\
\dot{n}_{R}^{\pm} &=& P_{R}^{\pm}-\Gamma_{R}^{\pm} n_{R}^{\pm} - 
Q_{R}^{\pm}(n_{R}^{\pm}) N_{R}^{\pm} ,
\end{eqnarray}
\end{subequations}
where $P^{\pm}_{R,L}$ are the rates of exciton creation, usually induced by 
a suitable laser pumping \cite{BallariniNatCom2013}, 
$\Gamma^{\pm}_{L,R}$ are the decay rates of the excitons 
in the corresponding reservoir, and $Q^{\pm}_{L,R}(n^{\pm}_{L,R})$ are the rates 
of stimulated scattering of the reservoir excitons into the condensate
of $N_{L,R}^{\pm}$ polaritons, see Fig.~\ref{fig:fig1}. 
For simplicity, the scattering rate can be approximated 
by a linear function of the reservoir exciton population, 
$Q^{\pm}_{L,R}(n^{\pm}_{L,R}) \simeq q^{\pm}_{L,R} n^{\pm}_{L,R}$ ~\cite{LienPRB2015}.

The model can be further simplified if we assume sufficiently 
large reservoirs such that the exciton populations remain nearly 
constant in time, $\dot{n}_{L}^{\pm} \simeq 0$, obtaining $n_{L,R}^{\pm} \simeq 
\frac{P_{L,R}^{\pm}}{\Gamma_{L,R}^{\pm} + q_{L,R}^{\pm} N_{L,R}^{\pm}}$.
Assuming also that $q_{L,R}^{\pm} N_{L,R}^{\pm}  \ll \Gamma_{L,R}^{\pm}$, 
i.e. the reservoirs are only weakly depleted by the coupling to 
the polariton condensates, as compared to their strong pumping and decay, 
we have $n_{L,R}^{\pm} = P_{L,R}^{\pm}/\Gamma_{L,R}^{\pm}$. Upon substitution 
into Eqs.~(\ref{eqs:EPs}), the polariton gain/loss coefficients 
become 
\begin{equation}
\gamma_{L,R}^{\pm} = \frac{1}{2}[q_{L,R}^{\pm} P_{L,R}^{\pm}/\Gamma_{L,R}^{\pm} - \kappa_{L,R}^{\pm}] .
\end{equation}
Hence, for given polariton decay rates $\kappa_{L,R}^{\pm}$ -- determined 
by photon losses from the cavity and the spontaneous decay (recombination) of the excitons --
and the stimulated scattering $q_{L,R}^{\pm}$ and decay $\Gamma_{L,R}^{\pm}$ rates of the 
reservoir excitons, the polariton gain/loss coefficients $\gamma_{L,R}^{\pm}$ can be 
precisely tuned by the laser pumping rates $P_{L,R}^{\pm}$. 
In turn, the intensity and spatial distribution of the left- and right-circularly polarized radiation
leaking from the cavity is directly proportional to the population $N_{L,R}^{\pm}$ of the corresponding 
polariton components and can thus serve for their continuous monitoring \cite{Ohadi2017}.

We note that interactions of the polaritons with the reservoir excitons 
\cite{AmoPhysique2016,EstrechoPRB2019} can strongly affect the polariton dynamics, leading 
to, e.g., instabilities \cite{DeveaudPhysique2016}, induce the polariton energy shifts 
and even be used to optically engineer the polariton landscape \cite{FerrierPRL2011,Ohadi2017}.
But under our assumption of the nearly-constant populations of the exciton reservoirs,
their interaction with the polaritons merely leads to a modification of the tight-binding 
trapping potentials and thereby to constant energy shifts which can be absorbed into $\varepsilon_{L,R}^{\pm}$.

A one-dimensional system, to be invariant under $\mathcal{PT}$ transformation
\cite{Bender1999,Bender2007,ElGanainy2018}, should be confined in a complex 
potential which has reflection symmetric and reflection antisymmetric real 
and imaginary parts, respectively. 
For the coupled-mode equations of the polariton system considered 
here, these requirements translate to the conditions
\begin{subequations} 
\label{eqs:PTcond}
\begin{align}
\varepsilon^{\pm}_{L} &= \varepsilon^{\pm}_{R} , \label{eq:PTcond1} \\
\gamma_{L}^{\pm} &= - \gamma_{R}^{\pm} \label{eq:PTcond2}.
\end{align}
\end{subequations}
We assume that the system is initially prepared with equal number 
of particles in each well, $N_{L}^{-} + N_{L}^{+} = N_{R}^{-} + N_{R}^{+} = N$. 
We then switch on the tunnel coupling $J$ and increase 
the reservoir pumping rate near the left well and reduce it 
near the right, in order to balance gain and loss according 
to the $\mathcal{PT}$ symmetry condition (\ref{eq:PTcond2}).  
The dynamics of the system is governed by the equations 
\begin{subequations} 
\label{eqs:eps}
\begin{eqnarray}
i \dot{\psi}_{L }^{\pm} &=& g_s |\psi_{L}^{\pm}|^{2} \psi_{L}^{\pm} 
+ g_{c} |\psi_{L}^{\mp}|^{2} \psi_{L}^{\pm}  - J \psi_{R}^{\pm} 
+ i \gamma \psi_{L}^{\pm} , \quad \\
i \dot{\psi}_{R }^{\pm} &=& g_s |\psi_{R}^{\pm}|^{2} \psi_{R}^{\pm} 
+ g_c |\psi_{R}^{\mp}|^{2} \psi_{R}^{\pm}  - J \psi_{L}^{\pm} 
- i \gamma \psi_{R}^{\pm},
\end{eqnarray}
\end{subequations}
where we set $\hbar = 1$ and the zero-point energies 
$\varepsilon^{\pm}_{L}=\varepsilon^{\pm}_{R}=0$,
and assumed the balanced gain/loss coefficients equal for both species,
$\gamma = \gamma_{L}^{\pm} = - \gamma_{R}^{\pm}$. 

\section{$\mathcal{PT}$ symmetry breaking and fixed points}
\label{sec:III}

Let us first review the properties of a linear system with $g_{s}=g_{c}=0$
\cite{Guo2009,Ruter2010}. 
The ($+$) and ($-$) polaritons decouple from each other and their equations 
become equivalent. The $\mathcal{PT}$ symmetric and broken phases can be 
determined from the eigenvalues of the corresponding Hamiltonian matrix, 
$\Lambda_{\pm}= \pm \sqrt{J^{2}-\gamma^2}$.
For $\gamma \leq J$, we have the $\mathcal{PT}$ symmetric phase 
and the system exhibits pseudo-Hermitian dynamics:
each polariton component coherently oscillates between the two wells,
\begin{gather}
\begin{bmatrix} 
\psi_{L}(t) \\
\psi_{R}(t) 
\end{bmatrix} = 
U
\begin{bmatrix} 
\psi_{L}(0) \\
\psi_{R}(0) 
\end{bmatrix} , \nonumber \\
U =   
\begin{bmatrix}  
\cos (\Omega t) + \frac{\gamma}{\Omega} \sin (\Omega t) & i \frac{J}{\Omega} \sin (\Omega t)  \\
i \frac{J}{\Omega} \sin (\Omega t) & \cos (\Omega t) - \frac{\gamma}{\Omega} \sin (\Omega t)  
\end{bmatrix} , \label{eq:PTlineval}
\end{gather}
with the effective Rabi frequency $\Omega = (\Lambda_+ -\Lambda_-)/2 = \sqrt{J^{2}-\gamma^2}$.
When $\gamma > J$, the eigenvalues $\Lambda_{\pm}$ become imaginary, 
the $\mathcal{PT}$ symmetry breaks, and the population of the particles 
in both wells will diverge exponentially.    

In the presence of interactions, $g_{s},g_{c} \neq 0$, the $\mathcal{PT}$ phase 
diagram cannot be determined from the eigenvalues of non-linear Hamiltonian,
but the relevant information can be extracted from the fixed points and their 
stability properties \cite{Li2011,DuanmuPhilTrans2013}. 
Assuming $N_{L}^{+} + N_{R}^{+} = N_{L}^{-} + N_{R}^{-} = N$, 
we can recast Eqs.~(\ref{eqs:eps}) in terms of the polariton population 
imbalances $z^{\pm}= (N_{L}^{\pm} - N_{R}^{\pm})/N$ and phase differences 
$\Phi^{\pm}=\phi_{R}^{\pm}-\phi_{L}^{\pm}$ as  
\begin{subequations} 
\label{eqs:zphi}
\begin{align}
 \dot{z}^{\pm} &= - 2\sqrt{1-(z^{\pm})^2} \sin\Phi^{\pm}+2\frac{\gamma}{J}, \\
\dot{\Phi}^{\pm} &= \frac{g_{s}}{J} z^{\pm} + \frac{g_{c}}{J} z^{\mp}+2\frac{z^{\pm}}{\sqrt{1-(z^{\pm})^2}}\cos\Phi^{\pm}, 
\end{align}
\end{subequations}
where we use the dimensionless time $\tau = tJ$. 
The fixed points correspond to the static solutions of these equations, 
$\dot{z}^{\pm}=\dot{\Phi}^{\pm}=0$. By linearizing the above equations 
in the vicinity of a fixed point, we find the stability eigenvalues 
$\lambda_i$ which determine whether the fixed point is 
(i) \textit{stable} if all eigenvalues have negative real parts,
(ii) \textit{unstable} if one or more eigenvalues have positive real part, and 
(iii) \textit{elliptic} if all eigenvalues are imaginary. 
The dependence of the stability eigenvalues on $\gamma$ thus 
serves as the phase diagram of the nonlinear polariton system, since it 
shows when the system diverges and when it follows Hermitian-like dynamics. 

\begin{figure}[t]
\begin{center}
\includegraphics[width=\columnwidth]{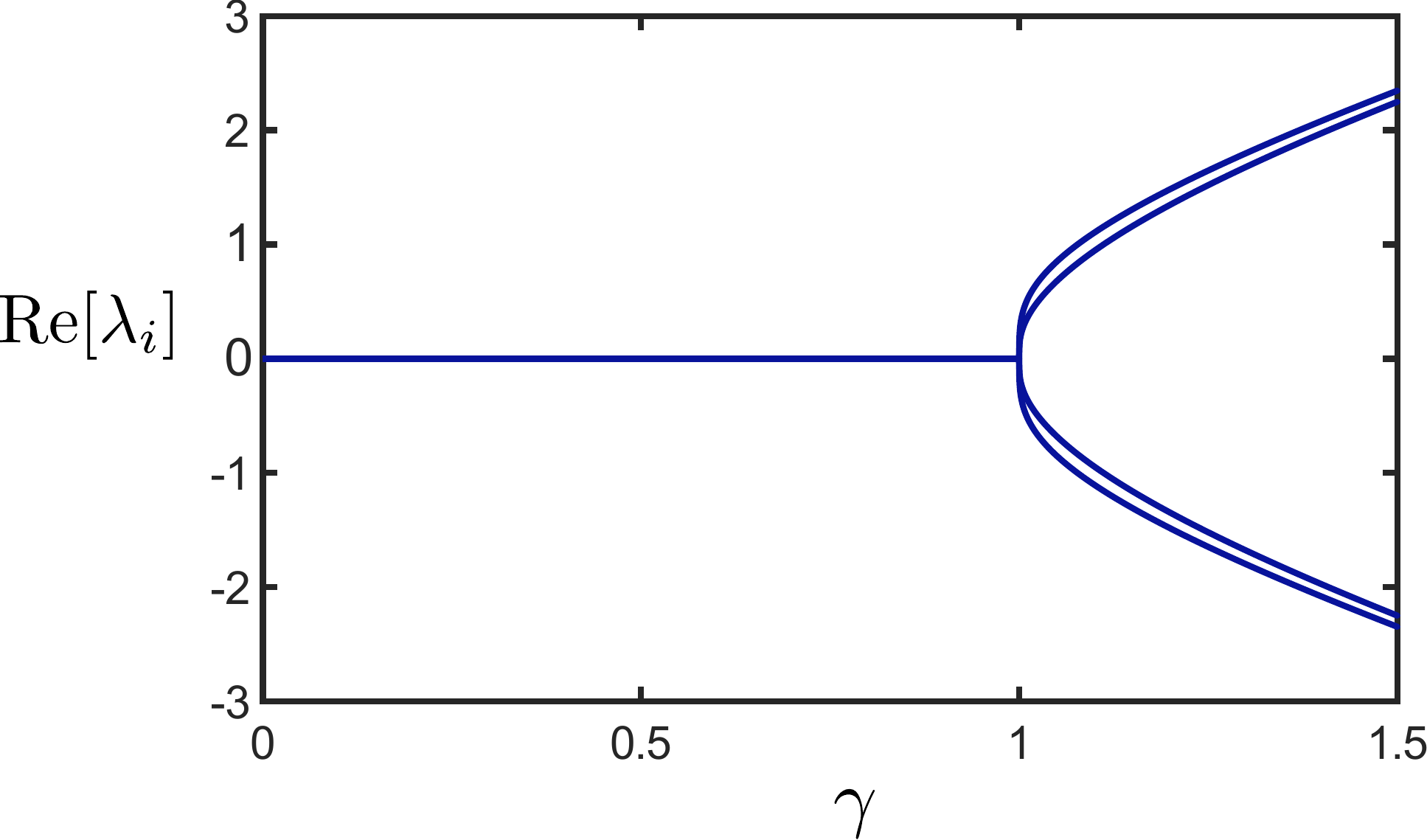}
\end{center}
\caption{Real part of the stability eigenvalues $\lambda_{i}$ of 
Eq.~(\ref{eq:lambdai}) around the fixed point (\ref{eq:fixed_point_1})
as a function of $\gamma$, for $g_{s} N =J$ and $g_c N =0.9J$.
Both axes are in units of $J$.}
\label{fig:fig2}
\end{figure}

A fixed point of the system corresponds to the trivial solution
\begin{equation}
z_{1}^{\pm} = 0 , \quad \Phi_{1}^{\pm}=\pm\arcsin\left(\frac{\gamma}{J}\right),
\label{eq:fixed_point_1}
\end{equation}
with equal population of the $(+)$ and $(-)$ polaritons in each well. 
The analytic expression for the four stability eigenvalues is
\begin{equation}
\lambda_{i} = \pm \frac{\sqrt{-4(J^{2}-\gamma^{2})-2(g_{s} \pm g_{c})\sqrt{(J^{2}-\gamma^{2})}}}{J} . \label{eq:lambdai}
\end{equation}
In Fig.~\ref{fig:fig2} we show the real parts of the stability eigenvalues 
$\lambda_{i}$ as a function of $\gamma$, for the system prepared in the vicinity
of the fixed point~(\ref{eq:fixed_point_1}). We observe only elliptic fixed
points, $\mathrm{Re}[\lambda_{i}]=0$, where any small deviation leads to
oscillations around the fixed point; and unstable fixed points, 
$\mathrm{Re}[\lambda_{i}] \neq 0$, in the vicinity of which the system diverges.
Equations~(\ref{eq:lambdai}) indicate that as long as $g_{s}>g_{c}$ the 
bifurcation to the unstable eigenvalues occurs when $\gamma$ reaches 
a threshold value equal to $J$. For the values of $g_{c}$ sufficiently larger 
than $g_{s}$ the bifurcation point is shifted to smaller values of $\gamma$.  

We have examined the possibility of stable fixed points 
with unequal populations of the two wells, corresponding to, 
e.g., self-trapping of polariton population on one well 
\cite{Albiez2005,AbbarchiNatPhys2013}. 
Our analytical and numerical results do not support the existence 
of such fixed points, as they break the $\mathcal{PT}$ symmetric 
phase of the system \cite{Ramezani2010,GraefeJPA2012}.

\section{Dynamics of the system}
\label{sec:IV}

\begin{figure}[t]
\begin{center}
\includegraphics[width=\columnwidth]{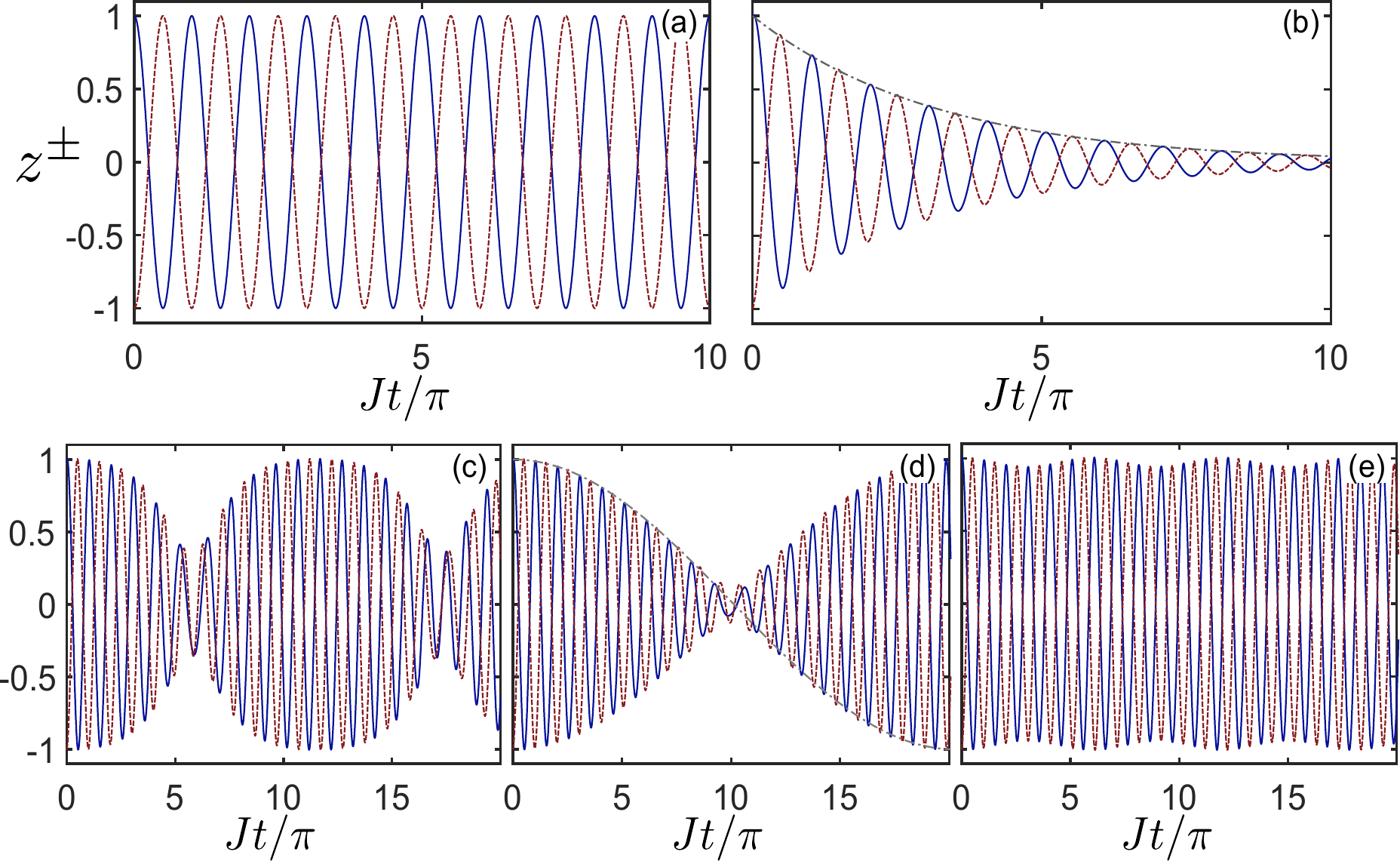}
\end{center}
\caption{Dynamics of the population imbalances $z^{\pm}$ of two polariton 
species in a double well system, for
(a) Hermitian case $\gamma=0$,
(b) non-Hermitian case, with the net gain $\gamma_L=0.1J$ on the left well and
net loss $\gamma_R=-0.2J$ on the right well, and 
(c,d,e) $\mathcal{PT}$ symmetric cases with balanced gain/loss 
$\gamma_L = - \gamma_R= \gamma=0.1J$ on the two wells, and the interaction strengths
(c) $g_{s}N= J$, $g_{c}N=0.8 J$, (d) $g_{s}N = g_{c}N=gN=J$, and (e) $g_{s}N= J$, $g_{c}N=1.2 J$.
The dash-dotted (gray) line for the amplitude modulation in (b) is $\exp[ (\gamma_L + \gamma_R)t]$, 
and in (d) is $\cos(\frac{\gamma g N}{2J} t)$.
Note the different time scales in (a,b) and (c,d,e).}
\label{fig:fig3}
\end{figure}

Before we examine the dynamics of our two-component $\mathcal{PT}$ symmetric 
system, it is instructive to consider its Hermitian version with $\gamma=0$. 
We assume an initial configuration with all ($+$) particles in the left well
and all ($-$) particles in the right. Figure~\ref{fig:fig3}(a) illustrates  
the corresponding dynamics of population imbalances 
$z^{\pm}= (N_{L}^{\pm} - N_{R}^{\pm})/N$. We observe that the population 
of each species exhibits Rabi-like oscillations between the two wells 
with frequency $J$.

The exciton polariton system is inherently dissipative, which necessitates
continuous pumping of polaritons from the exciton reservoirs. 
In Fig.~\ref{fig:fig3}(b) we illustrate the dynamics of the system
with the net loss from the right well, $\gamma_R=-0.2J$, being larger 
than the net gain on the left well, $\gamma_L=0.1J$. As 
expected, the polariton populations decay to zero with rate 
$\gamma_L + \gamma_R$, as do the population imbalances $z^{\pm}$ 
normalized to the initial particle number $N$. In the opposite case 
of gain being larger than loss, the polariton populations would diverge. 

The Hermitian-like dynamics can be achieved with balanced loss and gain,
$\gamma_L = -\gamma_R = \gamma$, as per Eqs.~(\ref{eqs:PTcond}) corresponding 
to the $\mathcal{PT}$ symmetric system. Note that the Hermitian system with
$\gamma_{L,R}=0$ can be considered as a special case of the $\mathcal{PT}$
symmetric system with zero imaginary part of the potential. 
In Fig.~\ref{fig:fig3}(c,d,e) we illustrate the dynamics of the 
interacting system.  
In the case of $g_s \geq g_c$, the dynamics is characterized by slow 
modulation of the amplitude of the Rabi-like oscillations between the two wells
(see Fig.~\ref{fig:fig3}(c)).
These modulations are not sinusoidal, exhibiting plateaus of maximal amplitude, 
followed by rapid collapse and revival of the oscillations. In the case 
of $g_s = g_c = g$, the modulation amplitude becomes harmonic, with the period
well approximated by $T = \frac{4\pi J}{\gamma g N}$ (see Fig.~\ref{fig:fig3}(d)). 
But for moderate interaction strengths $g_s < g_c$, 
the dynamics is nearly indistinguishable from that of the Hermitian system
(see Fig.~\ref{fig:fig3}(e)).

\section{Simulating two-qubit \textsc{swap} gate}
\label{sec:swap}

The two-species polariton system in a $\mathcal{PT}$ symmetric double well 
can be formally mapped onto a system of two qubits, or spin-1/2 particles, 
coupled via exchange interaction. This is an interesting analogy as it links 
an essentially classical system described by coupled-mode equations 
to a quantum system of coupled spins. 
Here we consider a \textsc{swap} gate to simulate state transfer between two qubits.

The complete basis for a system of two qubits (spins) consists of the four 
states $\{ \ket{\downarrow \downarrow}, \ket{\downarrow \uparrow}, 
\ket{\uparrow \downarrow}, \ket{\uparrow \uparrow} \}$,
and an arbitrary two-qubit state $\ket{\Psi}$ can be expanded as
\begin{equation}
\label{eq:2qubitstate} 
\ket{\Psi} = C_{\downarrow \downarrow} \ket{ \downarrow \downarrow}
+ C_{\downarrow \uparrow} \ket{\downarrow \uparrow}
+ C_{\uparrow \downarrow} \ket{\uparrow \downarrow}
+ C_{\uparrow \uparrow}  \ket{\uparrow \uparrow}.
\end{equation}
We may associate the ($+$) component of the polariton in each well 
$\psi_{L,R}^{+}$ with the amplitude of the $\ket{\uparrow }$ state 
of the spin and the ($-$) component $\psi_{L,R}^{-}$ with the amplitude 
of the $\ket{\downarrow}$ state of the spin. 
Then, the superposition amplitudes in Eq.~(\ref{eq:2qubitstate}) are given by 
$C_{\downarrow \downarrow} = c_{L}^{-} c_{R}^{-}$,
$C_{\downarrow \uparrow} = c_{L}^{-} c_{R}^{+}$,  
$C_{\uparrow \downarrow} = c_{L}^{+} c_{R}^{-}$,
$C_{\uparrow \uparrow} = c_{L}^{+} c_{R}^{+}$, where
$c_{L}^{\pm} \equiv \frac{\psi_{L}^{\pm}} 
{\sqrt{|\psi_{L}^{+}|^{2} + |\psi_{L}^{-}|^{2}}}$ and similarly for
$c_{R}^{\pm} \equiv \frac{\psi_{R}^{\pm}} 
{\sqrt{|\psi_{R}^{+}|^{2} + |\psi_{R}^{-}|^{2}}}$

Our aim is to realize the \textsc{swap} gate between the two spins representing
qubits. Consider the four input states 
$\ket{\downarrow \downarrow}, \ket{\downarrow \uparrow}, 
\ket{\uparrow \downarrow}, \ket{\uparrow \uparrow}$, 
which can cast as vectors 
\begin{equation}
u_{\downarrow \downarrow}=\begin{bmatrix} 
1  \\
0  \\
0  \\
0 
\end{bmatrix} , \; 
u_{\downarrow \uparrow}=\begin{bmatrix} 
0  \\
1  \\
0  \\
0 
\end{bmatrix} , \; 
u_{\uparrow \downarrow}=\begin{bmatrix} 
0  \\
0  \\
1  \\
0 
\end{bmatrix} , \; 
u_{\uparrow \uparrow}=\begin{bmatrix} 
0 \\
0  \\
0  \\
1
\end{bmatrix} .
\end{equation}
In the simplest, non-interacting and Hermitian case, 
$g_{s}=g_{c}=\gamma=0$ and $J \neq 0$, the dynamics of the system is 
analytically solvable and for each input state the time-dependent 
amplitudes of the output state (\ref{eq:2qubitstate}) are given in 
Table~\ref{table1}. We observe periodic oscillations of the excitation 
between the two spins, or qubits, with frequency $J$. Choosing for the 
interaction time $t$ half the period of the oscillations, $tJ = \pi/2$,
yields the transformation
\begin{equation}
\label{swap} U_{\textsc{swap}}=-\begin{bmatrix} 
1 & 0 & 0 & 0  \\
0 & 0 & 1 & 0 \\
0 & 1 & 0 & 0 \\
0 & 0 & 0 & 1
\end{bmatrix} ,
\end{equation}
which is precisely the \textsc{swap} gate (with an overall minus sign). 
For comparison, we recall the dynamics of the coupled spin system
governed by the spin-exchange ($XY$) Hamiltonian 
$H = - J \hat{\sigma}_L^+ \hat{\sigma}_R^- + \mathrm{H.c.}$, 
with $\hat{\sigma}^+ = \ket{\uparrow}\bra{\downarrow}$ and 
$\hat{\sigma}^- = \ket{\downarrow}\bra{\uparrow}$, for which the initial 
states $\ket{\downarrow \downarrow}$ and $\ket{\uparrow \uparrow}$ remain 
unchanged, while the other two basis states evolve as
\begin{eqnarray*}
\ket{\downarrow \uparrow} & \to &\cos(Jt) \ket{\downarrow \uparrow}
 + i \sin(Jt) \ket{\uparrow \downarrow} , \\
\ket{\uparrow \downarrow} & \to &\cos(Jt) \ket{\uparrow \downarrow}
 + i \sin(Jt) \ket{\downarrow \uparrow} . 
\end{eqnarray*}
The precise dynamics of quantum spin system differs from that in Table I, 
but at time $tJ=\frac{\pi}{2}$ we obtain the equivalent $i$\textsc{swap} gate
\begin{equation}
\label{iswap} U_{i\textsc{swap}}= \begin{bmatrix} 
1 & 0 & 0 & 0  \\
0 & 0 & i & 0 \\
0 & i & 0 & 0 \\
0 & 0 & 0 & 1
\end{bmatrix} .
\end{equation}

\begin{table}[t]
\centering 
\setlength\tabcolsep{5pt}
\begin{tabular}{|c |c |c |c |c|} 
\hline
$~$ & $u_{\downarrow \downarrow} $ & $u_{\downarrow \uparrow}$ & $u_{\uparrow \downarrow}$ & $u_{\uparrow \uparrow}$    \\ [1.5ex] 
\hline  
$C_{\downarrow \downarrow} $  & $e^{2iJt}$ &$ i \sin(Jt) \cos(Jt)$ & $ i \sin(Jt) \cos(Jt)$   & 0   \\ [2ex]  \hline 
$C_{\downarrow \uparrow}$  & $0$ & $\cos^2(Jt)$ & $-\sin^2(Jt)$ & $0$  \\  [2ex] \hline 
$C_{\uparrow \downarrow}$& $0$ & $-\sin^2(Jt)$ & $\cos^2(Jt)$ & $0$ \\  [2ex] \hline  
$C_{\uparrow \uparrow}$ & 0 & $ i \sin(Jt) \cos(Jt)$ & $ i \sin(Jt) \cos(Jt)$  & $e^{2iJt}$  \\   [2ex]
\hline 
\end{tabular}
\caption{ \label{table1} The time dependent amplitudes  
$C_{\downarrow \downarrow},C_{\downarrow \uparrow},C_{\uparrow \downarrow},C_{\uparrow \uparrow}$
of the two-qubit state in Eq.~(\ref{eq:2qubitstate}), for four different 
inputs $u_{\downarrow \downarrow},u_{\downarrow \uparrow},u_{\uparrow \downarrow},u_{\uparrow \uparrow}$, 
for a non-interacting, Hermitian system, $g_{s}=g_{c}=\gamma=0$. 
The table represents the elements of the transformation matrix $U$.}
\end{table}

Consider now the dynamics of the interacting, non-Hermitian $\mathcal{PT}$ 
symmetric system. For each input state $u_i$ at $t=0$, the output state
$\ket{\Psi_o}$ at time $t >0$ is   
\begin{equation}
\label{eq:2qubitstatei}
\ket{\Psi_o^{(i)}}= C^{(i)}_{\downarrow \downarrow} \ket{\downarrow \downarrow} + 
C^{(i)}_{\downarrow \uparrow} \ket{\downarrow \uparrow} + 
C^{(i)}_{\uparrow \downarrow} \ket{\uparrow \downarrow} + 
C^{(i)}_{\uparrow \uparrow} \ket{\uparrow \uparrow} .
\end{equation}
We can construct the transformation matrix $U$ as follows: 
For each input state $u_{i}$, the amplitudes of the output state
$C^{(i)}_{o}$ ($i,o \in \{ \downarrow \downarrow, \downarrow \uparrow,
\uparrow \downarrow,\uparrow \uparrow \}$) constitute the $i$th column of $U$.  
The complete matrix is then obtained by evaluating the dynamics of the system
for the four input basis states. 

We characterize gate transformation using the fidelity~\cite{Molmer2007},
\begin{equation}
\label{fidelity} 
F=\frac{1}{n(n+1)} \left[ \textrm{Tr}(MM^{\dagger}) + |\textrm{Tr}(M)|^2 \right],
\end{equation}
where $n = 4$ is the size of the Hilbert space (matrix dimension), and  
$M=U_{0}^{\dagger}U$ with $U_{0}$ the unitary transformation corresponding 
to the  desired quantum gate, e.g., $U_{\textsc{swap}}$, and $U$ is the actual 
transformation that we obtain from our simulations. In other words, for 
any initial state $\ket{\Psi}$, the desired final state is $U_{0} \ket{\Psi}$ 
while the actual state is $U\ket{\Psi}$, and Eq.~(\ref{fidelity}) gives 
the fidelity of the transformation averaged over all the input states. 

The linear $\mathcal{PT}$ case, $g_s=g_c=0$, is analytically solvable
via Eq.~(\ref{eq:PTlineval}). At half-period $tJ=\frac{\pi}{2}$ we then obtain 
the fidelity $F=0.992$ for $\gamma=0.1J$. The small reduction of fidelity is 
due to incomplete population transfer, 
$\min\limits_{t} |C_{\downarrow \uparrow (\uparrow \downarrow)}| \simeq \frac{\gamma^2}{J^2-\gamma^2}$,
and a slight shift (stretching) of the oscillation frequency by $\gamma$, i.e., 
the effective Rabi frequency is $\Omega \equiv \sqrt{J^2-\gamma^2} \simeq J( 1 - \gamma^2/2J^2)$. 
For larger $\gamma=0.3J$, the fidelity at $tJ=\frac{\pi}{2}$ is further reduced
to $F=0.935$. The obtained fidelities can be slightly improved by choosing 
appropriately delayed \textsc{swap} time $tJ=\frac{\pi}{2 -\gamma^2/J^2}$
to compensate for the reduction of the effective Rabi frequency of the oscillations.

\begin{figure}[t]
\centerline{\includegraphics[width=1\columnwidth]{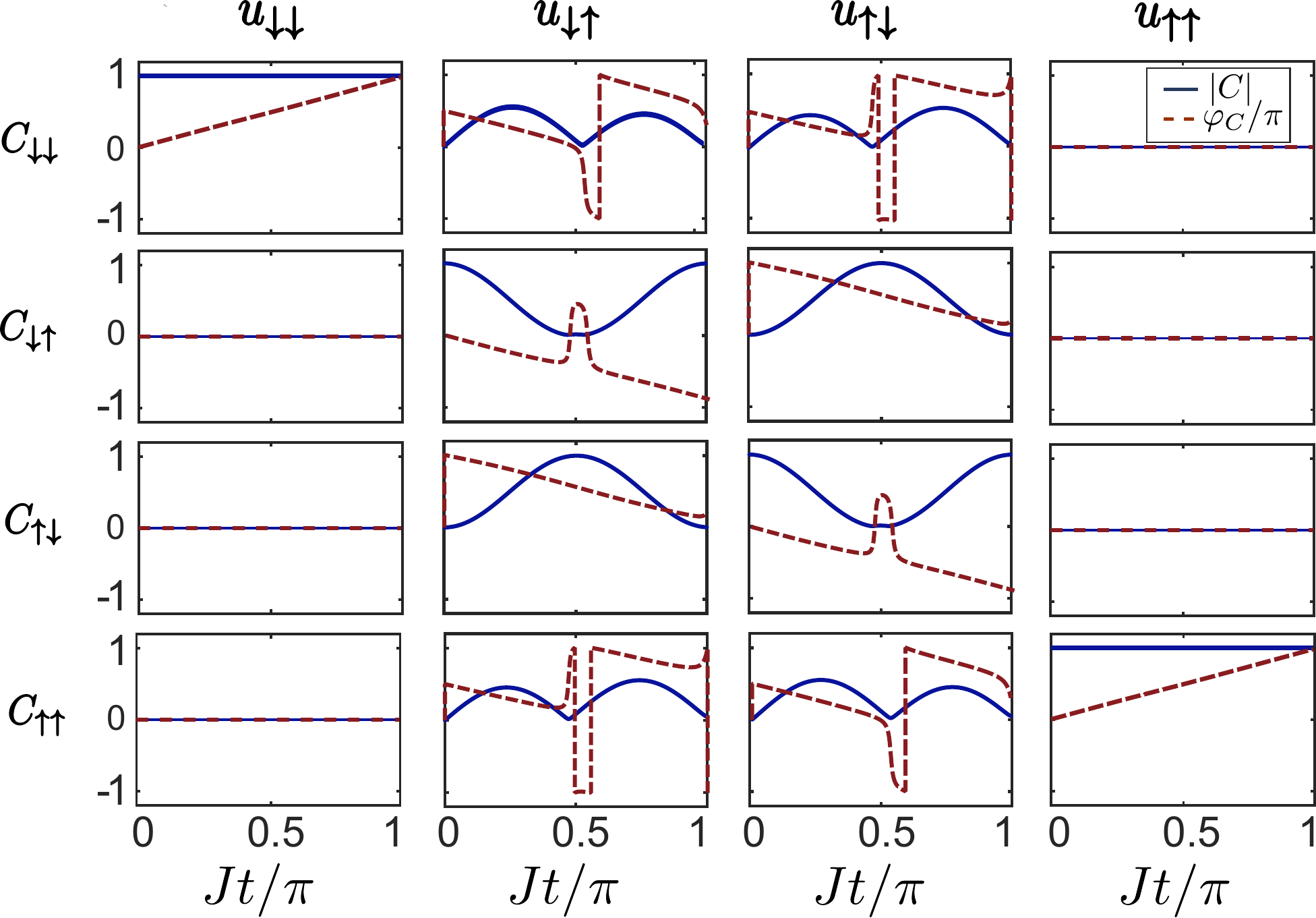}}
\caption{Time dependence of the complex amplitudes $C_{o}$ of the output 
state $\ket{\Psi_o}$ for each input state $u_{i}$, up to time $t=\pi/J$, 
for the system with $g_sN=J$, $g_cN=0.5J$ and $\gamma=0.1J$. 
In each panel, the blue (solid) line is the absolute value of the 
corresponding amplitude, $|C_o^{i}|$, and the red (dashed) line is the phase
$\phi_C = \arg (C_o^{i})$.}
\label{fig:fig4}
\end{figure}

We next consider the interacting system with equal self and cross-interaction 
strengths $g_sN=g_cN=J$ and $\gamma=0.1J$. This system is not analytically 
solvable and we numerically calculate the time-dependent amplitudes $C^{(i)}_{o}$.
We obtain fidelity $F=0.991$ of the \textsc{swap} gate at time $tJ=\frac{\pi}{2}$. 
Further increasing the loss/gain rate decreases the fidelity, 
e.g., for $\gamma=0.3J$ we obtain $F =0.922$, and for $\gamma=0.5J$ 
we obtain $F=0.799$, which can again be improved by using the 
optimal \textsc{swap} times $t = \frac{\pi}{2 \Omega}$. 

Finally, we examine the general case of unequal self- and cross-interaction 
strengths. Figure~\ref{fig:fig4} shows the evolution of the complex amplitudes 
$C^{(i)}_{o}$ for each input state $u_{i}$ for the system with 
$g_sN=J$, $g_cN=0.5J$ and $\gamma=0.1J$. Each panel of the figure is a direct 
visualization of the corresponding time-dependent matrix element of 
transformation $U$, up to time $tJ=\pi$ of one complete period of population 
oscillation for the corresponding linear, Hermitian system.
At half the period $tJ=\frac{\pi}{2}$, we obtain the  \textsc{swap} fidelity 
$F=0.982$, which is only slightly lower than in the previous cases. With 
increasing the difference between $g_{s}$ and $g_{c}$ the fidelity decreases, 
albeit rather slowly. For instance, for $g_sN=J$, $g_cN=0.1J$ the fidelity is 
$F=0.963$ for $\gamma=0.1J$ and $F=0.867$ for $\gamma=0.3J$.

\section{Conclusions}

To summarize, we have considered a two-species polariton mixture with self- and 
cross-interaction nonlinearities in a $\mathcal{PT}$ symmetric double well structure and examined
the static and dynamic properties of the system. We have shown that this essentially classical 
system described by the mean-field coupled-mode equations can nevertheless mimic quantum
dynamics of exchange-coupled spins implementing quantum gates. 

Our work can be relevant to analog simulations of few-body quantum systems with coupled 
exciton-polaritons in lattice potentials \cite{AmoPhysique2016,KlembtAPL2017,WhittakerPRL2018,OhadiPRB2018,PanPRB2019}.
Thus, one-dimensional polariton lattices can be used to study the dynamics of a single spin excitation,
e.g. for the characterization of faithful state or excitation transfer in spin chains 
with engineered couplings and tunable disorder \cite{Petrosyan2010,Nikolopoulos2014}. 
Furthermore, the quantum problem of two interacting particles or spin excitations 
on a one-dimensional lattice can be mapped onto an appropriately designed two-dimensional lattice of polaritons,
similarly to the simulations of the one-dimensional two-particle Hubbard model with two-dimensional arrays 
of coupled optical waveguides \cite{Krimer2011,Corrielli2013,Rai2015,Mukherjee2016}. 
Finally, tunable polariton lattices can serve as versatile simulators of classical spin-lattice models, 
such as Ising or $XY$ \cite{LagoudakisNJP2017,BerloffNatMater2017,KalininNJP2018}, 
which in turn can tackle certain NP-hard optimization problems.

\acknowledgments
We thank E. Paspalakis for fruitful discussions. 
This work was co-financed by Greece (General Secretariat for Research and Technology), 
and the European Union (European Regional Development Fund), in the framework of 
the bilateral Greek-Russian Science and Technology collaboration on Quantum Technologies 
(POLISIMULATOR project).



\begin{thebibliography}{99}

\bibitem{DengRevModPhys2010} H. Deng, H. Haug, and Y. Yamamoto, Rev. Mod. Phys. \textbf{82}, 1489 (2010).

\bibitem{CarusottoRevModPhys2010} I. Carusotto and C. Ciuti, Rev. Mod. Phys. \textbf{85}, 299 (2013).


\bibitem{YamaguchiPRL2013} M. Yamaguchi, K. Kamide, R. Nii, T. Ogawa, and Y. Yamamoto, Phys. Rev. Lett. \textbf{111}, 026404 (2013).

\bibitem{SchneiderNature2013} C. Schneider, A. Rahimi-Iman, N. Y. Kim, J. Fischer, I. G. Savenko, M. Amthor, M. Lermer, A. Wolf, L. Worschech, V. D. Kulakovskii, I. A. Shelykh, M. Kamp, S. Reitzenstein, A. Forchel, Y. Yamamoto, and S. Hofling, Nature \textbf{497}, 348 (2013).

\bibitem{KasprzakNature2006} J. Kasprzak, M. Richard, S. Kundermann, A. Baas, P. Jeambrun, J. Keeling, F. M.  Marchetti, M. H. Szymanska, R.  Andr\'e, J. L. Staehli, V. Savona, P. B. Littlewood, B. Deveaud, and L. S. Dang, Nature \textbf{443}, 409 (2006).

\bibitem{BaliliScience2007} R. Balili, V. Hartwell, D. Snoke, L. Pfeiffer, K. West, Science \textbf{316}, 1007 (2007).

\bibitem{LerarioNatPhys2017} G. Lerario, A. Fieramosca, F. Barachati, D. Ballarini, K. S. Daskalakis, L. Dominici, M. De Giorgi, S. A. Maier, G. Gigli, S. K\'ena-Cohen, and D. Sanvitto, Nature Phys. \textbf{13}, 837 (2017).

\bibitem{WoutersPRB2008} M. Wouters, I. Carusotto, and C. Ciuti, Phys. Rev. B 77, 115340 (2008).

\bibitem{TralleroPRB2014} C. Trallero-Giner, M. V. Durnev, Y. Nunez Fernandez, M. I. Vasilevskiy,V. Lopez-Richard, and A. Kavokin, Phys. Rev. B \textbf{89}, 205317 (2014).


\bibitem{WoutersPRL2007} M. Wouters and I. Carusotto, Phys. Rev. Lett. \textbf{99}, 140402 (2007).

\bibitem{LagoudakisPRL2010} K. G. Lagoudakis, B. Pietka, M. Wouters, R. Andr\'e, and B. Deveaud-Pl\'edran, Phys. Rev. Lett. \textbf{105}, 120403 (2010).

\bibitem{RahmaniSciRep2016} A. Rahmani and F. P. Laussy, Sci. Rep. \textbf{6}, 28930 (2016). 

\bibitem{AbbarchiNatPhys2013} M. Abbarchi, A. Amo, V. G. Sala, D. D. Solnyshkov, H. Flayac, L. Ferrier, I. Sagnes, E. Galopin, A. Lemaître, G. Malpuech, and J. Bloch, Nature Phys. \textbf{9}, 275 (2013).

\bibitem{ZhangOQE2017} Rui Zhang, Tao Wang, Zhong Chang Zhuo, Huifang Zhang, and Xue Mei Su, Opt. Quantum Electron. \textbf{49}, 205 (2017). 

\bibitem{AmoPhysique2016} A. Amo, J. Bloch, C. R.Physique \textbf{17}, 934 (2016).

\bibitem{KlembtAPL2017} S. Klembt, T. H. Harder, O. A. Egorov, K. Winkler, H. Suchomel, J. Beierlein, M. Emmerling, C. Schneider, and S. H\"ofling, Appl. Phys. Lett. \textbf{111}, 231102 (2017).

\bibitem{WhittakerPRL2018} C.E. Whittaker, E. Cancellieri, P.M. Walker, D.R. Gulevich, H. Schomerus, D. Vaitiekus, B. Royall, D.M. Whittaker, E. Clarke, I.V. Iorsh, I.A. Shelykh, M.S. Skolnick, and D.N. Krizhanovskii, Phys. Rev. Lett. \textbf{120}, 097401  (2018).

\bibitem{OhadiPRB2018} H. Ohadi, Y. del Valle-Inclan Redondo, A.J. Ramsay, Z. Hatzopoulos, T.C.H. Liew, P.R. Eastham, P.G. Savvidis, J.J. Baumberg, Phys. Rev. B \textbf{97}, 195109 (2018).

\bibitem{PanPRB2019} H. Pan, K. Winkler, M. Powlowski, M. Xie, A. Schade, M. Emmerling, M. Kamp, S. Klembt, C. Schneider, T. Byrnes, S. H\"ofling, and N. Y. Kim, Phys. Rev. B \textbf{99}, 045302 (2019).

\bibitem{BallariniNatCom2013} D. Ballarini, M. De Giorgi, E. Cancellieri, R. Houdr\'e, E. Giacobino, R. Cingolani, A. Bramati, G. Gigli \& D. Sanvitto, Nature Commun. \textbf{4},  1778 (2013).



\bibitem{Bender1999} C. Bender, S. Boettcher, P. Meisinger, J. Math. Phys. \textbf{40}, 2201 (1999).

\bibitem{Bender2007} C.M. Bender, Rep. Prog. Phys. \textbf{70}, 947 (2007).

\bibitem{ElGanainy2018}
R. El-Ganainy, K. G. Makris, M. Khajavikhan, Z. H. Musslimani, S. Rotter, and D. N. Christodoulides, 
Nature Phys. \textbf{14}, 11 (2018).

\bibitem{Bender2002} C.M. Bender, D.C. Brody, and H.F. Jones, Phys. Rev. Lett. {\bf 89}, 270401 (2002).

\bibitem{Bender2004} C.M. Bender, D.C. Brody, and H.F. Jones, Phys. Rev. D {\bf 70}, 025001 (2004).

\bibitem{Ganainy2007} R. El-Ganainy, K.G. Makris, D.N. Christodoulides and Z.H. Musslimani, Opt. Lett. \textbf{32}, 2632 (2007).

\bibitem{Musslimani2008a} Z.H. Musslimani, K.G. Makris, R. El-Ganainy, and D.N. Christodoulides, Phys. Rev. Lett. {\bf 100}, 030402 (2008). 

\bibitem{Peng2014} B. Peng, S. K. \"Ozdemir, F. Lei, F. Monifi,	M. Gianfreda, G. L. Long, S. Fan, F. Nori, C. M. Bender and L. Yang, Nat. Phys. \textbf{10}, 394 (2014).

\bibitem{Zhu2014} X. Zhu, H. Ramezani, C. Shi, J. Zhu, and X. Zhang, Phys. Rev. X {\bf 4}, 031042 (2014).

\bibitem{Fleury2015} R. Fleury,	D. Sounas, and A. Al\'u, Nat. Comm. {\bf 6}, 5906 (2015).

\bibitem{Guo2009} A. Guo, G. J. Salamo, D. Duchesne, R. Morandotti, M. Volatier-Ravat, V. Aimez, G. A. Siviloglou, and D. N. Christodoulides, Phys. Rev. Lett. {\bf 103}, 093902 (2009).

\bibitem{Ruter2010} C. E. R\"uter, K. G. Makris, R. El-Ganainy, D. N. Christodoulides, M. Segev and D. Kip, Nat. Phys. {\bf 6}, 192 (2010).

\bibitem{Makris2010} K. G. Makris, R. El-Ganainy, D. N. Christodoulides and Z. H. Musslimani,  Phys. Rev. A {\bf 81}, 063807 (2010).

\bibitem{Makris2008} K. G. Makris, R. El-Ganainy, D. N. Christodoulides and Z. H. Musslimani, Phys. Rev. Lett. {\bf 100}, 103904 (2008).

\bibitem{Ramezani2010} H. Ramezani, T. Kottos, R. El-Ganainy, and D. N. Christodoulides, Phys. Rev. A \textbf{82}, 043803 (2010).

\bibitem{Ambroise2012} J. D'Ambroise, P. G. Kevrekidis, and S. Lepri, J. Phys. A \textbf{45}, 444012 (2012).

\bibitem{Yang2015} F. Yang and Z. L. Mei, Sci. Rep. \textbf{5}, 14981 (2015).

\bibitem{Kottos2010} T. Kottos, Nat. Phys. \textbf{6}, 166 (2010).

\bibitem{Li2011} K. Li and P. G. Kevrekidis, Phys. Rev. E \textbf{83}, 066608 (2011).


\bibitem{Li2013} K. Li, P. G. Kevrekidis, D. J. Frantzeskakis, C. E. R\"uter and D. Kip, J. Phys. A \textbf{46}, 375304 (2013).

\bibitem{Zezyulin2012} D. A. Zezyulin and V. V. Konotop, Phys. Rev. Lett. \textbf{108}, 213906 (2012).



\bibitem{LienPRB2015} J.-Y. Lien, Y.eh-N. Chen, N. Ishida, H.-B. Chen, C.-C. Hwang, and F. Nori, Phys. Rev. B \textbf{91}, 024511 (2015).

\bibitem{GaoNature2015} T. Gao, E. Estrecho, K. Y. Bliokh, T. C. H. Liew, M. D. Fraser, S. Brodbeck, M. Kamp, C. Schneider, S. Höfling, Y. Yamamoto, F. Nori, Y. S. Kivshar, A. G. Truscott, R. G. Dall, and E. A. Ostrovskaya, Nature \textbf{526}, 554–558 (2015).

\bibitem{SatijaPRA2009} I. I. Satija, R. Balakrishnan, P. Naudus, J. Heward, M. Edwards, and C. W. Clark, Phys. Rev. A \textbf{79}, 033616 (2009).

\bibitem{ChestnovSciRep2016} I. Y. Chestnov, S. S. Demirchyan, A. P. Alodjants, Y. G. Rubo, and
A. V. Kavokin,  Sci. Rep. \textbf{6}, 19551 (2016). 

\bibitem{Ohadi2017}
H. Ohadi, A. J. Ramsay, H. Sigurdsson, Y. del Valle-Inclan Redondo, S. I. Tsintzos, Z. Hatzopoulos, T. C. H. Liew, I. A. Shelykh, Y. G. Rubo, P. G. Savvidis, and J. J. Baumberg,
Phys. Rev. Lett. \textbf{119}, 067401 (2017).


\bibitem{RaghavanPRA1999} S. Raghavan, A. Smerzi, S. Fantoni, and S. R. Shenoy, Phys. Rev. A \textbf{593}, 620 (1999).




\bibitem{EstrechoPRB2019} E. Estrecho, T. Gao, N. Bobrovska, D. Comber-Todd, M. D. Fraser, M. Steger, K. West, L. N. Pfeiffer, J. Levinsen, M. M. Parish, T. C. H. Liew, M. Matuszewski, D. W. Snoke, A. G. Truscott, and E. A. Ostrovskaya, Phys. Rev. B \textbf{100}, 035306 (2019). 

\bibitem{DeveaudPhysique2016} B. Deveaud C. R. Physique \textbf{17 }, 874 (2016).


\bibitem{FerrierPRL2011} L. Ferrier, E. Wertz, R. Johne, D. D. Solnyshkov, P. Senellart, I. Sagnes, A. Lemaître, G. Malpuech, and J. Bloch, Phys. Rev. Lett. \textbf{106}, 126401 (2011).

\bibitem{DuanmuPhilTrans2013} M. Duanmu, K. Li, R. L. Horne, P. G. Kevrekidis, N. Whitaker,  Phil. Trans. R. Soc. A \textbf{371}, 20120171 (2013).

\bibitem{Albiez2005}
M. Albiez, R. Gati, J. F\"olling, S. Hunsmann, M. Cristiani, and M. K. Oberthaler, Phys. Rev. Lett.  \textbf{95}, 010402 (2005).

\bibitem{GraefeJPA2012} E-M. Graefe, J. Phys. A: Math. Theor. \textbf{45}, 444015 (2012).

\bibitem{Molmer2007} 
L. H. Pedersen, N. M. M\o ller, and K. M\o lmer, Phys. Lett. A \textbf{367},  47 (2007).

\bibitem{Petrosyan2010}
D. Petrosyan, G. M. Nikolopoulos and P. Lambropoulos, Phys. Rev. A \textbf{81}, 042307 (2010)

\bibitem{Nikolopoulos2014}
G. M. Nikolopoulos and I. Jex (eds.),
{\it Quantum State Transfer and Network Engineering}, Quantum Science and Technology, (Springer, Heidelberg, 2014)

\bibitem{Krimer2011}
D. O. Krimer and R. Khomeriki,
Phys. Rev. A \textbf{84}, 041807(R) (2011).

\bibitem{Corrielli2013}
G. Corrielli, A. Crespi, G. Della Valle, S. Longhi, and R. Osellame, Nature Commun. \textbf{4}, 1555 (2013).

\bibitem{Rai2015}
A. Rai, C. Lee, C. Noh, and D. G. Angelakis,
Sci. Rep. \textbf{5} 8438 (2015).

\bibitem{Mukherjee2016}
S. Mukherjee, M. Valiente, N. Goldman, A. Spracklen, E. Andersson, P. \"O hberg, and R. R. Thomson,
Phys. Rev. A \textbf{94}, 053853 (2016).

\bibitem{LagoudakisNJP2017} 
P. G. Lagoudakis and N. G. Berloff, New J. Phys. \textbf{19}, 125008 (2017).

\bibitem{BerloffNatMater2017} 
N. G. Berloff, M. Silva, K. Kalinin, A. Askitopoulos, J. D. T\"opfer, P. Cilibrizzi, W. Langbein, and P. G. Lagoudakis, 
Nature Mater. \textbf{16}, 1120 (2017).

\bibitem{KalininNJP2018} 
K. P. Kalinin and N. G. Berloff,  New J. Phys. \textbf{20}, 113023 (2018).


\end{thebibliography}
\end{document}